\documentclass[epj]{webofc}
\usepackage[varg]{txfonts}   
\usepackage{array}
\usepackage{multirow}
\usepackage{wrapfig}

\newcolumntype{L}[1]{>{\raggedright\let\newline\\\arraybackslash\hspace{0pt}}m{#1}}
\newcolumntype{C}[1]{>{\centering\let\newline\\\arraybackslash\hspace{0pt}}m{#1}}
\newcolumntype{R}[1]{>{\raggedleft\let\newline\\\arraybackslash\hspace{0pt}}m{#1}}
\usepackage{braket}

\woctitle{MESON2014 - the 13$^\textrm{th}$ International Workshop on Meson Production, Properties and Interaction}
\begin{document}
\selectlanguage{english}
\title{Status of the measurement of $K_S \rightarrow \pi e \nu$ branching ratio and lepton charge
	asymmetry with the KLOE detector}

	\author{Daria Kamińska\inst{1}\fnsep\thanks{\email{daria.kaminska@uj.edu.pl}} 
			\\on behalf of the KLOE-2 collaboration
	}

\institute{Institute of Physics, Jagiellonian University,  ul. Reymonta 4, 30-059 Cracow, Poland 
}

\abstract{%
	We	present the current status of the analysis of about 1.7 billion $K_S K_L$
	pair	events collected at DA$\Phi$NE with the KLOE detector
	to	determine the branching ratio of $K_S \rightarrow \pi e \nu$ decay and the lepton
	charge	asymmetry. This sample is $\sim 4$ times larger in statistics
	than	the one used in the previous most precise result, from KLOE as well,
	allowing us to improve
	the	accuracy	on	the measurement and related tests of CPT symmetry and
	$\Delta S = \Delta Q$~rule.
}
\maketitle
\section{Introduction}
\label{intro}
	The $\mathcal{CPT}$ symmetry assumes invariance of physical laws under the combination of the
	symmetries such as charge conjugation~$(\mathcal{C})$, parity~$(\mathcal{P})$ and time
	reversal~$(\mathcal{T})$.
	One of possible ways to test violation of $\mathcal{CPT}$ symmetry and  basic assumptions of
	the Standard Model in the neutral kaon system is
	based~on the difference between charge asymmetries for short-lived kaon ($A_S$) and 
	for long-lived kaon ($A_L$).
	Presently this difference is compatible with zero within errors, which suggests
	conservation of $\mathcal{CPT}$ symmetry, however  the value of $A_L$~\cite{ktev_kl_charge_asymm} was determined with a precision more than
	two orders of magnitude better than $A_S$~\cite{kloe_final_semileptonic}.

 \section{Charge asymmetry and experimental verification}
 \label{sec-1}
	\begin{wrapfigure}{r}{0.4\textwidth}
		\vspace{-30pt}
		\centering
		\includegraphics[trim = 10mm 10mm 0mm 0mm, width=0.3\textwidth]{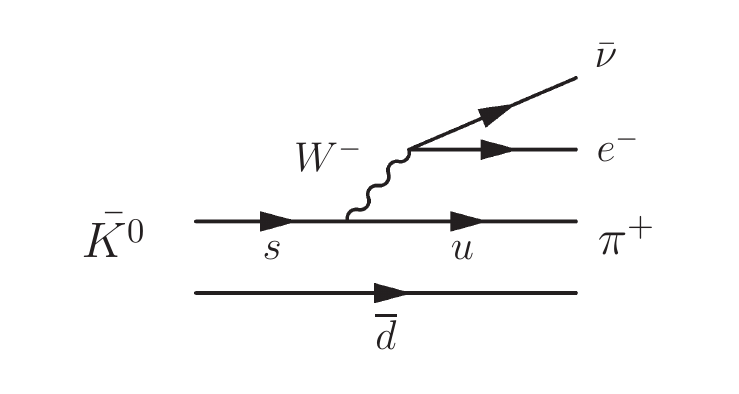}
		\includegraphics[trim = 10mm 10mm 0mm 0mm, width=0.3\textwidth]{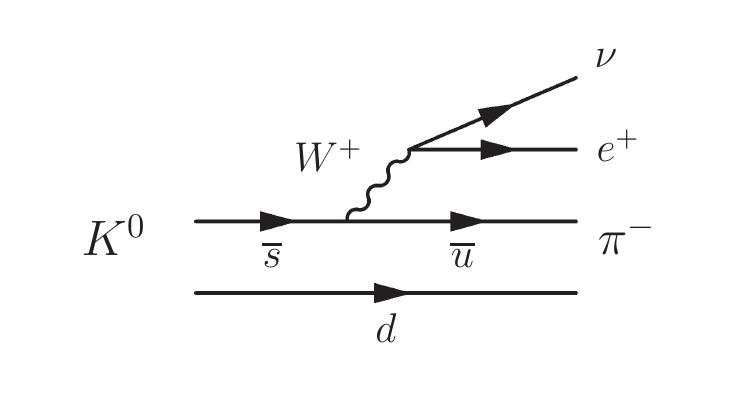}
		\caption{Feynman diagrams for $K^0$ and $\bar{K^0}$ semileptonic decays.}
		\label{decay}
		\vspace{-50pt}
	\end{wrapfigure}	
	According to the Standard Model, weak force is responsible 
	for semileptonic decay of $K^0$ or $\bar{K^0}$.
	This implies that only two of four possible $K^0$ or $\bar{K^0}$ semileptonic decays occur
	(Figure~\ref{decay}, Table~\ref{amplitudeParameters}) and 
	the change of strangeness $(\Delta S)$ entails  the corresponding change of electric
	charge~$(\Delta Q)$. This is the $\Delta S = \Delta Q$ rule.
	Semileptonic amplitudes can be parametrized as shown in~Table~\ref{amplitudeParameters}
	and connected to the conservation of discrete
	symmetries~(Table~\ref{connectionsAmplitudeSymmetries})~\cite{handbook_cp}.

		\begin{table}
   \begin{tabular}{|C{0.2\textwidth}||C{0.15\textwidth}|C{0.06\textwidth}C{0.21\textwidth}L{0.2\textwidth}|} \hline 
   {According to  the \newline Standard Model} &  Decay & \multicolumn{3}{c|}{Matrix element parametrization } \\  \hline \hline
    allowed &  $K^0 \rightarrow \pi^- e^+ \bar{\nu} $  &   
       $a + b$ & $= \bra{\pi^{-} e^{+} \nu } H_{weak}  \ket{K^{0}}$ &  $=\mbox{(if $\mathcal{CP}$)}= a^* + b^*  ,$
   \\             
     allowed & $  \bar{K^0} \rightarrow \pi^+ e^- \nu  $ &
      $ a^{*} - b^{*}$ & $= \bra{\pi^{+} e^{-} \bar{\nu} } H_{weak}  \ket{ \bar{K^{0}}} $  & $=\mbox{(if $\mathcal{CP}$)}= a+ b ,$ 
    \\
     not allowed  & $ K^0 \rightarrow \pi^+ e^- \nu $ &
     $ c + d $ & $= \bra{\pi^{+} e^{-} \bar{\nu} } H_{weak}  \ket{K^{0}}   $ & $=\mbox{(if $\mathcal{T}$)} = c^* - d^*,$
    \\ 
        not allowed &  $\bar{K^0} \rightarrow \pi^- e^+ \bar{\nu}$ &
      $c^{*} - d^{*}$ & $= \bra{\pi^{-} e^{+} \nu } H_{weak}  \ket{ \bar{K^{0}}}  $ & $=\mbox{(if $\mathcal{T}$)} = c + d.$ \\ \hline
   \end{tabular} 
			\caption{Relations between semileptonic decays of $K^0 (\bar{K^0})$, introduced parametrization
			and discrete symmetries.}
			\label{amplitudeParameters}
		\end{table}	

  \begin{table}
			\centering
    \begin{tabular}{|c|c||c|c|c|c|} \hline
    & & \multicolumn{4}{c|}{Conserved quantity  } \\ \cline{3-6}
    &   & $\mathcal{CP}$                & $\mathcal{T}$     & $\mathcal{CPT}$   & $\Delta S = \Delta
      Q$ \\ \hline \hline
    \multirow{4}{*}{  { \rotatebox{90}{ Parameter } }  } 
    & a & \textit{Im} = 0 & \textit{Im} = 0 &       &  \\
    & b & \textit{Re} = 0 & \textit{Im} = 0 &  = 0  &  \\
    & c & \textit{Im} = 0 & \textit{Im} = 0 &       & = 0 \\
    & d & \textit{Re} = 0 & \textit{Im} = 0 & = 0    & = 0 \\ \hline
    \end{tabular}
				\caption{Relations between discrete symmetries and semileptonic amplitudes.}
				\label{connectionsAmplitudeSymmetries}
   \end{table}

	\newpage \noindent
	Semileptonic amplitudes can be associated to the $K_S$ and $K_L$ semileptonic decay widths through
	the charge asymmetry:

		\begin{equation}
     \begin{aligned}
      A_{S,L} & = 
         \frac{\Gamma(K_{S,L} \rightarrow \pi^{-} e^{+} \nu) - \Gamma(K_{S,L}
         \rightarrow
         \pi^{+} e^{-}
         \bar{\nu})}{\Gamma(K_{S,L} \rightarrow \pi^{-} e^{+} \nu) + \Gamma(K_{S,L}
         \rightarrow
         \pi^{+} e^{-} \bar{\nu})}    \\
          & =  2 \left[ Re \left( \epsilon_{K} \right) \pm Re \left( \delta_{K}  \right) 
               + Re \left( \frac{b}{a} \right) \mp Re \left( \frac{d^{*}}{a} \right) \right] \nonumber \\
          &  \mbox{ if } \Delta Q = \Delta S  \nonumber \\
          & = 2 \left[ Re \left( \epsilon_{K} \right) \pm Re \left( \delta_{K}  \right) 
             + Re \left( \frac{b}{a} \right) \right] \nonumber \\
          &  \mbox{ if } \mathcal{CPT}  \mbox{ and } \Delta Q = \Delta S  \nonumber \\
          & = 2 \left[ Re \left( \epsilon_{K} \right)  \right].    \nonumber 
     \end{aligned}
    \end{equation}
	The charge asymmetry for $K_L$ was precisely determined from the KTeV experiment at
	Fermilab~\cite{ktev_kl_charge_asymm}:
    \begin{equation}
     A_{L} = (3.322 \pm 0.058_{stat} \pm 0.047_{syst}) \times 10^{-3},
    \end{equation}
	while the most precise measurement of $A_S$ was conducted by the KLOE collaboration~\cite{kloe_final_semileptonic}:
    \begin{equation}
     A_{S} = (1.5 \pm 9.6_{stat} \pm 2.9_{syst}) \times 10^{-3}.
    \end{equation}
	The obtained charge asymmetry for $K_S$ decays is consistent, within error, with the charge
	asymmetry for $K_L$ decays, which suggests conservation of $\mathcal{CPT}$ symmetry. 
	This result is dominated by the statistical uncertainty which is three times larger than the systematic
	contribution.

\newpage
\section{Measurement}
	\label{sec-2}
	The KLOE experiment is located at DA$\Phi$NE $e^+ e^-$ collider that works 
	at the center of mass energy of the $\phi$-meson mass ($\sqrt{s} = m_{\phi}$).  
	The KLOE detector was optimized for efficient detection of long-lived kaons. A $2$~m
	radius drift chamber allows to register around $40\%$
	of long-lived kaon decays inside the chamber while the rest reach the electromagnetic calorimeter. 
	Identification of events with long-lived kaon  
	ensures occurrence of short-lived kaon near the interaction point and vice versa.
	In order to improve signal over
	background ratio kinematic selection is applied. On remaining events the time-of-flight
	technique, which aims at rejecting background and identifying the final charge states ($\pi^+ e^-
	\bar{\nu}$ and $\pi^- e^+ \nu$), is
	used. Altogether about $10^5$ $K_S \rightarrow \pi e \nu$ decays were reconstructed, which will
	be used for the measurement of the charge asymmetry and branching ratio for $K_S$ semileptonic
	decays. The analysis is still in progress, nevertheless it shows potential of reaching a twice
	better statistical error determination based on four times larger data sample. Also, due to the upgrade
	of KLOE detector and DA$\Phi$NE collider, further reduction of systematical
	and statistical uncertainties are expected in the future~\cite{prospects_kloe}.

\begin{acknowledgement}
 We warmly thank our former KLOE colleagues for the access to the data collected during the KLOE data taking campaign.
 We thank the DA$\Phi$NE team for their efforts in maintaining low background running conditions and their collaboration during all data taking. We want to thank our technical staff: 
 G.F. Fortugno and F. Sborzacchi for their dedication in ensuring efficient operation of the KLOE computing facilities; 
 M. Anelli for his continuous attention to the gas system and detector safety; 
 A. Balla, M. Gatta, G. Corradi and G. Papalino for electronics maintenance; 
 M. Santoni, G. Paoluzzi and R. Rosellini for general detector support; 
 C. Piscitelli for his help during major maintenance periods. 
 This work was supported in part by the EU Integrated Infrastructure Initiative Hadron Physics Project under contract number RII3-CT- 2004-506078; by the European Commission under the 7th Framework Programme through the `Research Infrastructures' action of the `Capacities' Programme, Call: FP7-INFRASTRUCTURES-2008-1, Grant Agreement No. 227431; by the Polish National Science Centre through the Grants No. 
 DEC-2011/03/N/ST2/02641, 
 2011/01/D/ST2/00748,
 2011/03/N/ST2/02652,
 2013/08/M/ST2/00323,
 and by the Foundation for Polish Science through the MPD programme and the project HOMING PLUS BIS/2011-4/3.
\end{acknowledgement}
 %

\begin{thebibliography}{00}
 \bibitem{ktev_kl_charge_asymm} KTeV Collaboration, 
	Phys. Rev. Lett. \textbf{88}, 181601-181606  (2002)
 \bibitem{kloe_final_semileptonic} KLOE Collaboration, 
	Phys. Lett. \textbf{B636}, 173-182 (2006)
 \bibitem{handbook_cp} L. Maiani, G. Pancheri, N. Paver, \textit{The Second DA$\Phi$NE Physics	Handbook}
	(INFN-LNF, Frascati, 1995) 12-16
 \bibitem{prospects_kloe} KLOE-2 Collaboration, 
	Eur. Phys. J. \textbf{C68}, 619-681 (2010)
\end{thebibliography}
 %
 %

	\end{document}